\documentclass[final,5p,times,twocolumn]{elsarticle}

\usepackage{multirow}
\usepackage{tabu}
\usepackage{CJKutf8}
\usepackage[T1]{fontenc}
\usepackage[utf8]{inputenc}
\usepackage{graphics}
\usepackage{graphicx,color}
\usepackage{epsfig}
\usepackage{amsmath}
\usepackage{amssymb}

\usepackage{setspace}
\usepackage{hyperref}
\usepackage{cleveref}
\usepackage{underscore}
\usepackage[dvipsnames]{xcolor}
\usepackage{multirow, makecell}
\usepackage{colortbl}
\usepackage{dashrule}
\usepackage{ehhline}
\usepackage{csquotes}
\usepackage{natbib}
\biboptions{sort&compress}

\usepackage{array}
\usepackage{booktabs}
\usepackage{caption}
\makeatletter
\usepackage{colortbl}
\usepackage{hhline}

\journal{}
\begin{document}
	
	\begin{frontmatter}
		
		\title{Thermal Transport and Application Reassessment of ThSi$_2$N$_4$ Monolayer: From FET Channel to Thermoelectric Material}

		\author[1]{Maryam Mirzaei Farshmi}
		\author[1]{Seyedeh Ameneh Bahadori}
		\author[1]{Zahra Shomali\corref{cor1}}
		\address[1]{Department of Physics, Faculty of Basic Sciences, Tarbiat Modares University, Tehran 14115-175, Iran}
		\cortext[cor1]{Corresponding author, Tel.: +98(21) 82883147.}
		\ead{shomali@modares.ac.ir}
		
		\begin{abstract}
   The two-dimensional M$_2$Z$_4$ materials are proposed as suitable replacements for silicon channels in field-effect transistors (FETs). In the present work, the ThSi$_2$N$_4$ monolayer from the family, with the very appropriate electron mobility, is thermally investigated using the non-equilibrium Monte Carlo simulation of the phonon Boltzmann transport equation. The reliability of the MOSFET with the ThSi$_2$N$_4$ channel has been reassessed and determined to be low due to the high maximum temperature achieved. The phonon analysis is performed and reveals that the dominant contribution of fast and energetic LA and also slow and low energy ZA phonons alongside the minor participation of the TA phonons is responsible for the peak temperature rise reaching 800 K. This finding presents that the ThSi$_2$N$_4$ monolayer is not a good candidate for replacing as silicon channel but alternatively is capable of generating a significant temperature gradient, which makes it, a suitable candidate for using as a thermoelectric material in thermoelectric generators.
				
		\end{abstract}
		
	\end{frontmatter}
	
	\section{Introduction}
	Thermal management is pivotal in ensuring the reliability and efficiency of modern nanoelectronic devices, as miniaturization increases heat density and the risk of device failure \cite{Balachandra1991,Pop2006-2,Gong2015,Mahajan2002,Pop2005,Grosse2011,Moore2014}. The advent of two-dimensional (2D) semiconductors, such as graphene, MoS$_2$, and the newly MA$_2$Z$_4$ family including MoSi$_2$N$_4$, WSi$_2$N$_4$, and ThSi$_2$N$_4$, has introduced promising solutions for overcoming the limitations of silicon-based transistors in sub-5 nm scales \cite{Sheng2023,HLi2024,Kim2024,Hong2020,QWang2020,JYu2021,Shen2022,CLi2022,ChLu2023}. 
	
	Among these materials, the monolayer ThSi$_2$N$_4$ stands out due to its high thermal conductivity, suitable bandgap, ultra-high carrier mobility of 14384 cm$^2$V$^{-1}$s$^{-1}$, and structural stability, making it a potential candidate for using as a channel in next-generation field-effect transistors (FETs). On the other hand, making sure of the transistors' reliability, which is defined by their ability to withstand thermal and electrical stresses, is critical for their utilization in applications such as high-performance computing and energy-efficient electronics. Consequently, if a ThSi$_2$N$_4$ monolayer is an effective thermal management solution, it will result in mitigating the self-heating, which is a major cause of the performance degradation. Accordingly, the investigation of transient and steady state heat transport in 2D ThSi$_2$N$_4$ material is obligatory to check the monolayer's appropriateness considering the thermal issues. In Advanced simulation techniques, such as Monte Carlo methods and numerical solutions of the Boltzmann Transport Equation (BTE), have been employed to study phonon-mediated heat transport in 2D materials \cite{Shomali2018,Shomali2023,Bahadori2024,Bahadori2025}. Additionally, strain engineering, which has proven to be effective in enhancing the properties of materials \cite{Wang2019,Khengar2022}, also holds as a potential process for  optimizing the ThSi$_2$N$_4$ transistors. In the present study, the thermal management capabilities of ThSi$_2$N$_4$ monolayers and their heat dissipation characteristics in comparison with the other studied monolayers of MA$_2$Z$_4$ family; the MoSi$_2$N$_4$ and WSi$_2$N$_4$ 2D materials; is investigated. The study is performed using the non-equilibrium Monte Carlo simulations of the phonon Boltzmann transport equation. Our findings highlight that the ThSi$_2$N$_4$ monolayer achieves the higher peak temperature rise in response to the self-heating zone, compared to that of the 2D MoSi$_2$N$_4$ and WSi$_2$N$_4$ materials. This suggests that the ThSi$_2$N$_4$ transistor reliability is not very efficient. Hence, the monolayer is more suitable for use as the thermoelectric material \cite{Snyder2017,Hinterleitner2019,DLi2020,IJacob2024}. This paper follows the following structure. In Section \ref{Sec.2}, the geometry and the boundary conditions are introduced. The section \ref{Sec.3}, deals with mathematical modeling. The sections \ref{Sec.4} and \ref{Sec.5}, respectively, present the results and the conclusions.
	
	\section{Geometry and the boundary conditions}
	\label{Sec.2}
	
	The ThSi$_2$N$_4$ compound, typically crystallizes in a hexagonal crystal structure, exhibiting significant structural similarities to certain silicide and nitride compounds with a wurtzite-like arrangement. The reported lattice parameters for this compound are a=3.6$\AA$ and c=9.4$\AA$. Thorium atoms are located at the center of the structure, contributing to the stability of the lattice by forming strong metallic bonds. Additionally, silicon atoms are arranged in layered formations, creating Si–Si bonds. Nitrogen atoms also participate in the bonding process, enhancing the mechanical strength and thermal stability of the structure \cite{ChLu2023}. The monolayer ThSi$_2$N$_4$, which is studied here, has dimensions of 200 nm in both length, L$_x$, and width, L$_y$, with thickness of 0.735 nm. Also, the thickness of MoSi$_2$N$_4$ and WSi$_2$N$_4$ is, respectively, 0.679 nm and 0.702 nm, which shows that the family has nearly the same breadth. A uniform heat source of Q=10$^{12}$ W/m$^{3}$ is applied to the center region of the ThSi$_2$N$_4$ monolayer channel, specifically between x=45 nm and x=55 nm of the width, alongside the whole length. All boundaries, except the bottom boundary, are assumed to be adiabatic. The bottom boundary exchanges energy with the surroundings. Also, the initial temperature of the system is set to 299 Kelvin.
		
	\section{Mathematical Modeling}
	\label{Sec.3}
	
 The heat conduction at micro and nanoscale can be simulated using the atomistic or the phenomenological methods \cite{Shomali2012,Shomali20152,Shomali2016,Shomali2019,Sattler2022,Shomali2022}. In the present study, the problem is addressed using the non-equilibrium Monte Carlo method, which simulates the path of individual phonons based on their transmission and scattering behavior, effectively solving the phonon Boltzmann equation in a stochastic manner. In more detail, the phonon Boltzmann transport equation (BTE), models heat conduction at the microscopic level, and Monte Carlo simulation provides a powerful method for numerically solving it by tracking a large number of phonons as they move and scatter, and by considering the complex physics involved \cite{Mazumder2001,Mittal2011,Shomali2017,2Shomali2017}. Monte Carlo (MC) simulation is a stochastic approach that models individual phonons as particles to represent the entire distribution statistically. In this method, a large number of phonons are generated with the randomly initialized positions, directions, frequencies, $\omega$, and polarizations and in continuous, the number of phonons and the density of states are calculated. Finally, finding the energy in each cell, the pseudo temperature will be obtained \cite{Shomali2018}. Knowing the relation of frequency versus wave vector, the phonon dispersion curve, is essential for solving the BTE. In the case of our studied thermally isotropic 2D materials, the phonon dispersion curve has six branches including the three acoustic and three optical ones. In most two-dimensional cases, acoustic phonons play a much more prominent role in heat transport. Therefore, in the present work, only the acoustic phonon branches are taken into account. The acoustic phonon dispersion curves under consideration are considered as quadratic polynomials of the form $\omega$$_b$=c$_{b}$k$^{2}$+v$_ {b}$k \cite{Shomali2018}. In the table \ref{Tab1-tab1}, the fitting coefficients for the fitted quadratic equations are listed. 
	
	\begin{table*}[htbp]
		\caption{The coefficients of the fitted quadratic formula, ($\omega_b$=c$_b$k$^2$+v$_b$k), for three low-dimensional materials of ThSi$_2$N$_4$,  MoSi$_2$N$_4$, and WSi$_2$N$_4$.}
		\label{Tab1-tab1}
		\vspace{-0.5cm}
		\hskip -2cm
		\centering
		\begin{center}
			\begin{small}
				\begin{tabular}{|p{2.2cm}|p{1.6cm}|p{1.6cm}|p{1.6cm}|p{0.6cm}|p{0.6cm}|p{0.7cm}|}
						\hline
						
						2D Material &  c$_{LA}$(m$^2$/s) & c$_{TA}$(m$^2$/s) & c$_{ZA}$(m$^2$/s) & v$_{LA}$ (m/s) & v$_{TA}$ (m/s) & v$_{ZA}$ (m/s)\\
						\hline
						ThSi$_2$N$_4$ & {-3.952$\times$10$^{-8}$} & {-3.508$\times$10$^{-8}$}& {-3.079$\times$10$^{-8}$} & {896.9} & {747.7} & {653.17}  \\
						\hline
						WSi$_2$N$_4$ &{-3.392$\times$10$^{-7}$} & {-2.256$\times$10$^{-7}$} &  {-1.4$\times$10$^{-7}$} & {8188} & {5816} & {4886} \\ 
						\hline
						MoSi$_2$N$_4$ & {-4.252$\times$10$^{-7}$} & {-2.562$\times$10$^{-7}$} & {-1.259$\times$10$^{-7}$}& {10031} & {6982} & {5660} \\
						\hline
						
					\end{tabular}
				\end{small}
			\end{center}
		\end{table*}
		
		Moreover, the general form of the Boltzmann transport equation is as follows,
		
		\begin{equation}
			\label{eq1}
			\frac{\partial f(\mathbf{r},\mathbf{k},t)}{\partial t} + \mathbf{v_g}(\mathbf{k}) \cdot \nabla f(\mathbf{r},\mathbf{k},t)= \left( \frac{\partial f}{\partial t} \right)_{\text{scatt}}.
		\end{equation}
	
On the left side of the Eq. \ref{eq1}, the parameter $\mathbf{v_g}(\mathbf{k}$), is the phonon group velocity and $n(\mathbf{k},t)$, is the phonon distribution function, which depends on the phonon wave vector, $\mathbf{k}$, and the time t. The right side of the Eq. \ref{eq1}, often referred to as the collision term that determines the phonons tend to be driven towards the equilibrium with a characteristic time of $\tau$. Each phonon is drifted with the group velocity v$_g$ in a time step $\Delta$t, and scatters from the other phonons with the lifetime of $\tau$. The phonon-phonon scattering rate of Umklapp in each branch is given by the relation $\tau^{-1}_{U}(\omega)=\frac{\hbar \gamma^{2}}{\bar{M} \Theta v^{2} \omega^{2} T e^{-\Theta_b/3T}}$. In the mentioned formula, the parameter v is the speed of sound in each branch and $\bar{M}$ is the average atomic mass. The values of the parameter $\Theta$, $\bar{M}$ as well as Gruneissen, $\gamma$ are given in Table \ref{Tab2-tab2} for three monolayers of MA$_2$Z$_4$ family.
		
		\begin{table*}[htbp]
	\caption{The parameters involved in phonon-phonon scatterings for the monolayers ThSi$_2$N$_4$, WSi$_2$N$_4$ and MoSi$_2$N$_4$.  \newline}
	\label{Tab2-tab2}
	\vspace{-0.5cm}
	\hskip -2cm
	\centering
	\begin{small}
		
		\hspace*{-0.4cm}
		\begin{tabular}{|c|c|c|c|c|c|}
			\hline
			$Material$  & ThSi$_2$N$_4$ & WSi$_2$N$_4$ & MoSi$_2$N$_4$  \\
			\hline
			$\Theta$  & {740} & {433.33} & {335.3} \\
			\hline
			$\bar{M}$ (e$^{-27}$ kg) & {49.15}  & {42.29} & {29.74} \\
			\hline
			$\gamma$ &  {0.4975} & {0.4916} & {0.42} \\
			\hline
		\end{tabular}
	\end{small}
\end{table*}
		
		\section{Numerical considerations}
		\label{Sec.4}
 A ThSi$_2$N$_4$ monolayer is discretized to a uniform mesh of size 200$\times$200 in the XY plane. Also, the heat source region is located in the center of the channel producing the heat flux of Q=10$^{19}$ W/m$^3$. The frequency range for all three acoustic branches is divided into 1000 parts. For every 1000 frequency interval, the phonon relaxation time, which is obtained by inverting the scattering rate, is calculated for each branch. Simultaneously, the velocity of all phonons for each frequency is obtained. Then, by dividing the grid size by the all velocities calculated, the phonon travel time for each interval is determined. Finally, the minimum value of the calculated phonon travel time and relaxation time is selected as the time step.Finally, according to the calculations, the time steps of 3.47$\times$10$^{-13}$, 2.07$\times$10$^{-13}$ and 8.92$\times$10$^{-12}$ are obtained for MoSi$_2$N$_4$,  WSi$_2$N$_4$  and ThSi$_2$N$_4$ respectively. 
	
		\section{Results and Discussions}
		\label{Sec.5}
		
\begin{figure*}
			\vspace{0cm}
			\centering
		\includegraphics[width=1.5\columnwidth]{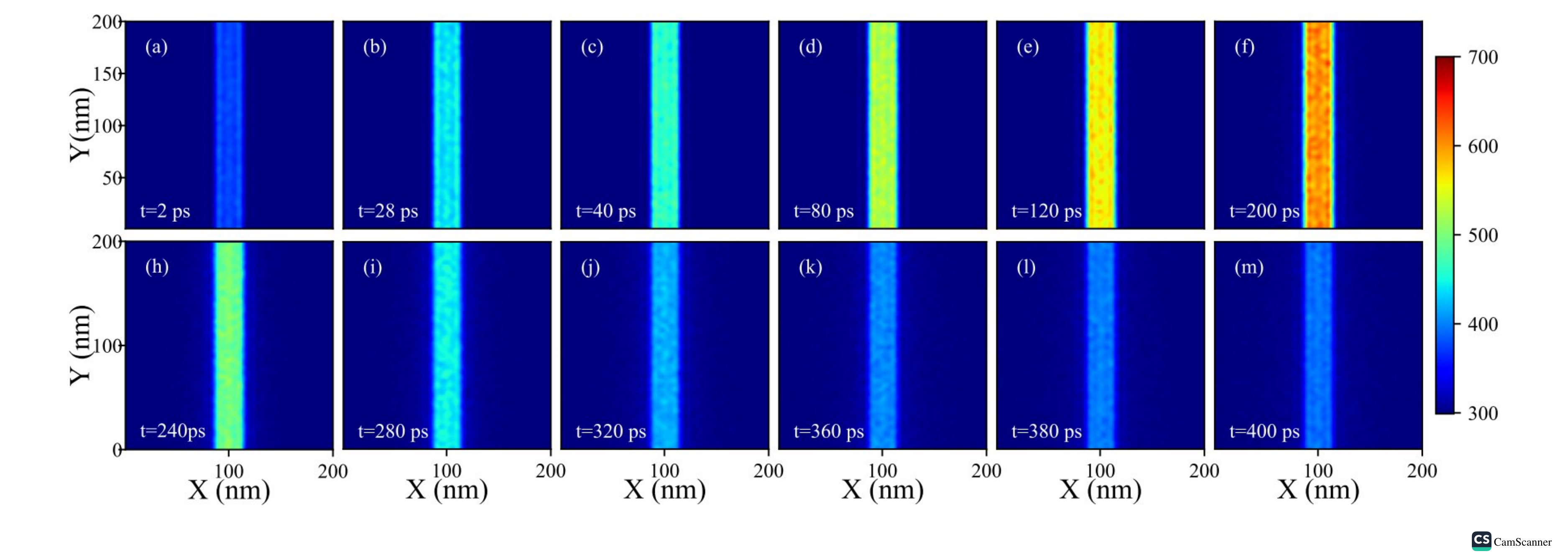}
			\caption{\label{prof2d} Hotspot temperature distribution on the XY plane during the heating and cooling process at time steps (a) 2, (b) 28, (c) 40, (d) 80, (e) 120, and (f) 160 ps for the 2D material of ThSi$_2$N$_4$.}
		\end{figure*}

\begin{figure}
			\vspace{0cm}
			\centering
			\includegraphics[width=\columnwidth]{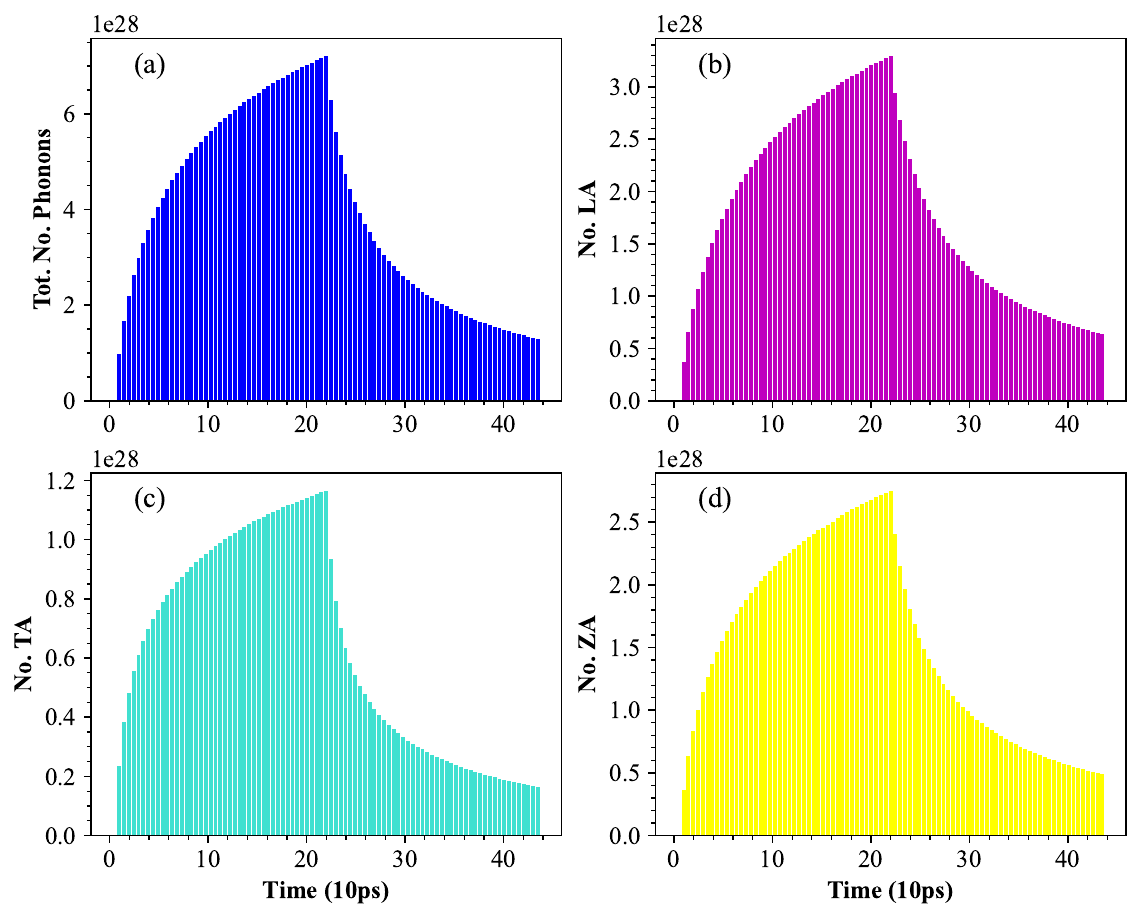}
			\caption{\label{phonons}The number of phonons per volume, for the 2D material of ThSi$_2$N$_4$ versus time, for (a) all the branches together, and (b) LA, (c) TA, and (d) ZA modes.}
		\end{figure}

During the simulation, the nanodevices according to the Joule-heating self-heating is heated up for the first 200 picoseconds, then for the next 200 ps of the simulation, the heat source is turned off and the channel is left to cool down.
As the Fig. \ref{prof2d} (a) presents, shortly after being influenced by the heat generation zone, the hot spots are formed in ThSi$_2$N$_4$ channel. During the heating process, according to the injection of the hot phonons to the channel from the gate, the overall energy of the monolayer transistor specially at the gate contact points is increased. As the Fig. \ref{phonons} shows, most of the phonons participating in the heat transport are the ones in ZA and LA modes. The ZA phonons are much slower than the transferring phonons in longitudinal or transverse modes. Hence, the nano-device turns out to be hot while the slow ZA phonons can not leave the hot zone easily. On the other hand, LA phonons are also dominant during the heating process. These phonons are faster than ZA and TA phonons, which reduces the likelihood of them being trapped in hotspots. However, they are highly energetic due to their high frequencies. As a result, even a small number of LA phonons trapped in a hotspot can carry significant energy, leading to a substantial temperature increase.

\begin{figure*}
			\vspace{0.0cm}
			\centering
		\includegraphics[width=1.5\columnwidth]{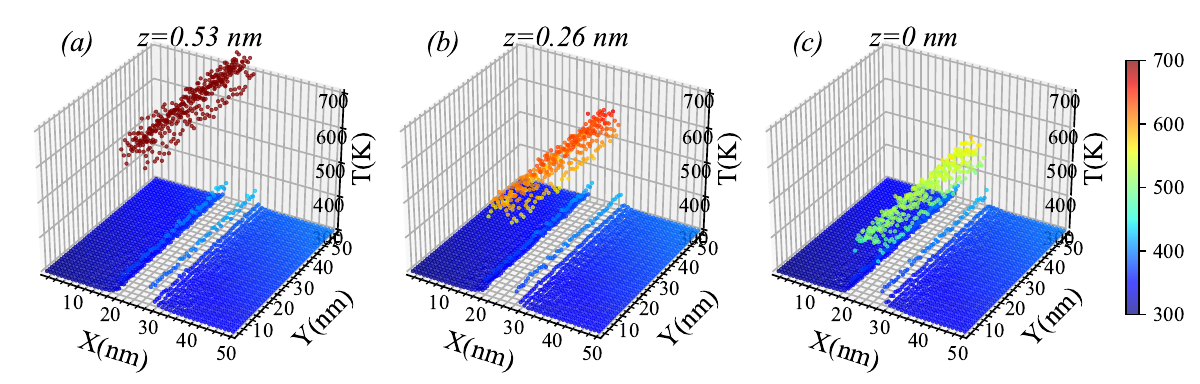}
			\caption{\label{prof3d} Temperature profile on the XY plane at t=200 ps, after the heating is switched off, for the ThSi$_2$N$_4$ monolayer at different depths: (a) z=thickness (top boundary), (b) z=half the thickness, and (c) z=0 (bottom boundary).}
	    	\end{figure*}

As mentioned, all boundaries are assumed to be adiabatic except the lower boundary, in which the device can exchange energy with the environment. As the monolayer has a small thickness, some phonons reach the lower boundary and leave the channel almost at the same time as they are created at the hotspot. For instance, in the Fig. \ref{prof3d}, the temperature behavior at t=200 ps, alongside the channel thickness is shown. At the interface of the channel with the gate, z=0.53 nm, as seen in Fig. \ref{prof3d} (a), the temperature is raised up to 800 K. While at z=0.26, the maximum temperature reached is 650 K. At the open bottom boundary, the hotspot is cooler due to its proximity to the external environment and the release of hot phonons, as shown in Fig. \ref{prof3d}(c).

		To analyze the peak temperature reached by ThSi$_2$N$_4$ during self-heating compared to other 2D materials in the MA$_2$Z$_4$ family, such as WSi$_2$N$_4$ and MoSi$_2$N$_4$, it is valuable to closely examine the types of phonons contributing to heat transfer. The status of the LA, TA, and ZA phonons involved in the heating and cooling process of the ThSi$_2$N$_4$ during the 400 picosecond simulation are shown in Fig. \ref{phonons}. Heat generation is modeled as a source that raises the temperature of the target zone by releasing phonons into it \cite{Wong2011,Wong2014}. Consequently, as long as the self-heating zone remains active, phonons emitted from the heat source are continuously added to those participating in thermal transport. Therefore, the phonon population reaches its peak at 200 picoseconds. Moreover, it is shown that the LA, ZA, and TA phonons, respectively, are the most dominant phonons, participating in thermal transport during the 400 ps of the simulation. The LA phonons are fast; however, they are also highly energetic. ZA phonons are slow and tend to localize in hotspots. Conversely, LA phonons escape more easily. On the other side, TA phonons exhibit behavior that is intermediate between LA and ZA phonons, which are faster than ZA but not as efficient as phonons in LA mode in energy transport. In other words, while LA phonons dominate heat conduction and ZA phonons are easily trapped, TA phonons play a balancing role in both mobility and energy transfer. The TA branch phonons, on one side, are fast to free themselves out of the hotspot, and on the other side, have frequencies that are comparable to the ZA phonons and not as large as that of the LA mode. Accordingly, the dominant co-existence of the much faster LA and TA phonons is one of the reasons for the lower maximum temperature. However, for the monolayer ThSi$_2$N$_4$, the dominant contributed phonons are the most energetic LA and least fast ZA phonons. Thus, the number of the TA phonons are in the minority and one expects the augmented maximum temperature. It is meaningful to highlight the results from the work by Shomali \cite{Shomali2023}, which investigated heat transport in two other members of the MSi$_2$N$_4$ family, the WSi$_2$N$_4$ and MoSi$_2$N$_4$ monolayers, for comparison with the thermal behavior of the ThSi$_2$N$_4$ monolayer studied here.It is confirmed that 23$\%$ to 46$\%$ of the contributing phonons in 2D WSi$_2$N$_4$ are of the TA type, while for MoSi$_2$N$_4$, this percentage ranges from 8$\%$ to 32$\%$. This phonon arrangement guarantees the lower peak temperature of the WSi$_2$N$_4$ relative to the MoSi$_2$N$_4$ monolayer.
		
\begin{figure}
			\vspace{0cm}
			\centering
			\includegraphics[width=\columnwidth]{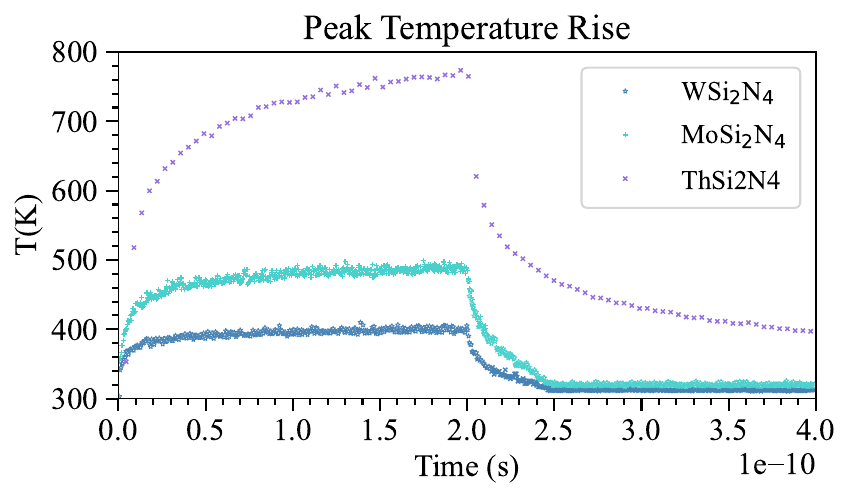}
			\caption{\label{peaktemp} The maximum temperature of the ThSi$_2$N$_4$ monolayer, compared to the other monolayers of the family, WSi$_2$N$_4$ and MoSi$_2$N$_4$.}
		\end{figure}

		As the phonon analysis for the monolayers from the MSi$_2$N$_4$ family is performed, one can investigate the reliability of the transistors with different kinds of MSi$_2$N$_4$ monolayers as channel, by calculating the maximum temperature they reach. The behavior of the maximum temperature versus time for three different monolayers of ThSi$_2$N$_4$, WSi$_2$N$_4$ and MoSi$_2$N$_4$, are demonstrated in Fig. \ref{peaktemp}. Among the single-layer 2D materials we have studied with MA$_2$Z$_4$ type, the 2D ThSi$_2$N$_4$ shows hot spots with highest maximum temperature. This can be justified as, the hot region paradoxically produces high-frequency phonons, but due to the dominant ZA phonons, alongside a minor presence of TA phonons, the heat dissipation is very low and more hot phonons are trapped in the hotspot.The phonon arrangement allows ThSi$_2$N$_4$ to reach a maximum temperature of 800 K, and after 200 ps of cooling, it stabilizes at 400 K, the peak temperature observed for the WSi$_2$N$_4$ monolayer. In contrast, the previously studied 2D materials WSi$_2$N$_4$ and MoSi$_2$N$_4$ exhibit maximum temperatures of 400 K and 500 K, respectively which are significantly lower than the peak temperature observed in the ThSi$_2$N$_4$ monolayer. As previously discussed, the improved thermal reliability associated with these lower maximum temperatures is attributed to the higher presence of transverse acoustic (TA) phonons, which is evident in the 2D WSi$_2$N$_4$ and MoSi$_2$N$_4$ materials \cite{Shomali2023}. 
		
		The ThSi$_2$N$_4$ monolayer with a high electron mobility alongside the two other members of the MSi$_2$N$_4$ family, 2D MoSi$_2$N$_4$, and WSi$_2$N$_4$, with excellent ambient stability, moderate band gap, considerable carrier mobility and favorable thermal conductivity, has been suggested as a potential channel for the field-effect transistors. In the present study, it is established that although the ThSi$_2$N$_4$ has appropriate electrical properties, but from thermal behavior point of view, having a very high maximum temperature, it lacks the necessary legitimacy for being used as the silicon channel replacement. As an alternative, being able to produce the high temperature gradient, the 2D ThSi$_2$N$_4$ material, is an adequate candidate for using as a thermoelectric material for the thermoelectric generators.
		
			\section{Conclusions}
			\label{Sec.6}
   In summary, the thermal transport properties of the ThSi$_2$N$_4$ monolayer have been thoroughly investigated using non-equilibrium Monte Carlo simulations of the phonon Boltzmann transport equation. Although, the high electron mobility of the ThSi$_2$N$_4$, at first place, introduces this material as potential candidate for silicon replacement, but the temperature rise reaching upto 800 K, cause doubts about such application. The analysis of the phonons participating in heat transport, reveals that the being the LA and ZA phonons as the dominant heat carriers, accounts for the augmented temperature hotspots. However, despite the limitation and concerns about MOSFET channel utilization of the ThSi$_2$N$_4$, this material presents promising characteristics for use as a thermoelectric material in thermoelectric generators. These findings highlight the importance of thoroughly understanding both the thermal and electrical properties of emerging 2D materials to determine where they can be used most effectively.
			


\begin{thebibliography} {10}  
	 \bibitem{Balachandra1991}
T.C. Balachandra, K. Ravindran, G.R. Nagabhushana, Numerical estimation of hotspot temperatures in electrodes subjected to pulsed electron beam heating in vacuum, Int Commun Heat Mass, 18(3), 397, 1991.

\bibitem{Pop2006-2}
E. Pop, S. Sinha, K.E. Goodson, Heat generation and transport in nanometer-scale transistors, Proceedings of the IEEE, 94(8), 1587, 2006.


\bibitem{Gong2015}
S. Gong, L. Chen, H. Feng, Z. Xie, F. Sun, Constructal optimization of cylindrical heat sources surrounded with a fin based on minimization of hot spot temperature, Int Commun Heat Mass, 68, 1, 2015.

\bibitem{Mahajan2002}
R. Mahajan, R. Nair, V. Wakharkar, J. Swan, J. Tang, G. Vandentop, Emerging Directions For Packaging Technologies. Intel Technology Journal. 6(2), 2002.

\bibitem{Pop2005}
E. Pop, R.W. Dutton, K.E. Goodson, Monte Carlo simulation of Joule heating in bulk and strained silicon, Appl. Phys. Lett. 86, 082101, 2005.

\bibitem{Grosse2011}
K.L. Grosse, M.H. Bae, F. Lian, E. Pop, W.P. King, Nanoscale Joule heating, Peltier cooling and current crowding at graphene-metal contacts, Nat. Nanotechnol. 6(5), 287, 2011.

\bibitem{Moore2014}
A.L. Moore, L. Shi, Emerging challenges and materials for thermal management of electronics, Mater. Today, 17(4), 163, 2014.

\bibitem{Sheng2023}
C. Sheng, X. Dong, Y. Zhu, X. Wang, X. Chen, Y. Xia, Z. Xu, P. Zhou, J. Wan, and W. Bao, Two‐Dimensional Semiconductors: From Device Processing to Circuit Integration, Adv. Funct. Mater. 33 (50), 2304778, 2023.

\bibitem{HLi2024}
H. Li, Q. Li, Y. Li, Z. Yang, R. Quhe, X. Sun, Y. Wang, L. Xu, L.M. Peng, H. Tian, and C. Qiu, Recent Experimental Breakthroughs on 2D Transistors: Approaching the Theoretical Limit. Advanced Functional Materials, 34 (38), 2402474, 2024.

\bibitem{Kim2024}
K.S. Kim, J. Kwon, H. Ryu, C. Kim, H. Kim, E.K. Lee, D. Lee, S. Seo, N.M. Han, J.M. Suh, and J. Kim, The future of two-dimensional semiconductors beyond Moore’s law, Nat. Nanotechnol., 19 (7), 895, 2024.

\bibitem{Hong2020}
Y. Hong, Z. Liu, L. Wang, T. Zhou, W. Ma, C. Xu, S. Feng, L. Chen, M. Chen, D. Sun, X. Chen, H. Cheng, and W. Ren, Chemical vapor deposition of layered two dimensional MoSi$_2$N$_4$ materials, Science, 369, 670, 2020.

\bibitem{QWang2020}
Q. Wang, L. Cao, S.-J. Liang, W. Wu, G. Wang, C. H. Lee, W. L. Ong, H. Y. Yang, L. K. Ang, S. A. Yang et al., Designing efficient metal contacts to two-dimensional semiconductors MoSi$_2$N$_4$ and WSi$_2$N$_4$ monolayers, arXiv preprint arXiv:2012.07465, 2020.

\bibitem{JYu2021}
J. Yu, J. Zhou, X. Wan, and Q. Li, High intrinsic lattice thermal conductivity in monolayer MoSi$_2$N$_4$, New Journal of Physics, 23(3), 033005, 2021.

\bibitem{Shen2022}
C. Shen, L. Wang, D. Wei, Y. Zhang, G. Qin, X.Q. Chen, and H. Zhang, Two-dimensional layered MSi$_2$N$_4$ (M= Mo, W) as promising thermal management materials: a comparative study, Physical Chemistry Chemical Physics, 24(5), 3086-3093, 2022.

\bibitem{CLi2022}
C. Li, and L. Cheng, Intrinsic electron transport in monolayer MoSi$_2$N$_4$ and WSi$_2$N$_4$, Journal of Applied Physics, 132(7), 075111, 2022.   

\bibitem{ChLu2023}
Ch. Lu, Ch. Cui, J. Zuo, H. Zhong, Sh. He, W. Dai, and X. Zhong, Monolayer ThSi$_2$N$_4$: An indirect-gap semiconductor with ultra-high carrier mobility, Phy. Rev. B., 108, 205427, 2023.

\bibitem{Shomali2023}
Z. Shomali, An investigation into the reliability of newly proposed MoSi$_2$N$_4$/WSi$_2$N$_4$ field effect transistor: A monte carlo study, Micro and nanostructures, 182(3), 207648, 2023.

\bibitem{Bahadori2024}
S. A. Bahadori, and Z. Shomali, Thermal transport in thermoelectric materials of SnSSe and SnS$_2$: A non-equilibrium Monte-Carlo simulation of Boltzmann transport equation, Case Stud. Therm. Eng., 57, 104377, 2024.

\bibitem{Bahadori2025}
S. A. Bahadori, Z. Shomali, and R. Asgari, Heat Dissipation and Thermoelectric Performance of InSe-Based Monolayers: A Monte Carlo Simulation Study, ArXiv (2025).

	\bibitem{Shomali2018}
Z. Shomali and R. Asgari, Eﬀects of low-dimensional material channels on energy consumption of nano-devices, Int. J. Heat Mass Transf., 94, 77,2018.

\bibitem{Wang2019}
Q. Wang, L. Han, L. Wu, T. Zhang, Sh. Li, and P. Lu, Strain Effect on Thermoelectric Performance of InSe Monolayer, Nanoscale Res Lett, 14, 287, 2019.

\bibitem{Khengar2022}
S. J. Khengar, P. R. Parmar, P. B. Thakor, Strain dependent structural and electronic properties of two-dimensional Janus In$_2$SeTe monolayer, Mater. Today, 67, Part 1, 2022.

\bibitem{Snyder2017}
G. J. Snyder and A. H. Snyder, Figure of Merit zT of a Thermoelectric Device Defined from Materials Properties, Energy \& Environmental Science, 10, 2280, 2017.

\bibitem{Hinterleitner2019}
B. Hinterleitner, I. Knapp, M. Poneder, Y. Shi, H. M$\ddot{u}$ller, G. Eguchi, C. Eisenmenger-Sittner, M. St$\ddot{o}$ger-Pollach, Y. Kakefuda, N. Kawamoto \emph{et al.}, Thermoelectric Performance of a Metastable Thin-Film Heusler Alloy, Nature, 576, 85, 2019

\bibitem{DLi2020}
D. Li, Y. Gong, Y. Chen, J. Lin, Q. Khan, Y. Zhang, Y. Li, H. Zhang, and H. Xie, Recent progress of two-dimensional thermoelectric materials, Nano-Micro Letters, 12, 1, 2020.

\bibitem{IJacob2024}
I. Jacob, R. Lamba, R. Kumar, and F. J.  Montero, Metaheuristic based single and multiobjective optimization of thermoelectric generator, Appl. Therm. Eng., 236, 121790, 2024.

 \bibitem{Shomali2012}
J. Ghazanfarian and Z. Shomali, Investigation of dual-phase-lag heat conduction model in a nanoscale metal-oxide-semiconductor field-effect transistor, Int. J. Heat Mass Transf., 55(21-22) (2012) 6231.

\bibitem{Shomali20152}
Z. Shomali, J. Ghazanfarian, A. Abbassi, Investigation of bulk/film temperature-dependent properties for highly non-linear DPL model in a nanoscale device: the case with high-k metal gate MOSFET, Superlattices Microstruct., 83 (2015) 699.

	\bibitem{Shomali2016}
Z. Shomali, A. Abbassi, and J. Ghazanfarian, Development of non-Fourier thermal attitude for three-dimensional and graphene-based MOS devices, Appl. Therm. Eng., 104, 616, 2016.

\bibitem{Shomali2019}
J. Ghazanfarian, Z. Shomali, and S. Xiong, 21st Century Nanoscience– A Handbook: Nanophysics Sourcebook (Volume One), Sattler, K. D. (Ed.), Chapter 4, CRC Press, 2019.		

\bibitem{Sattler2022}
K. D. Sattler, 21st Century Nanoscience: A Handbook (Ten-Volume Set), CRC Press, 2022.

\bibitem{Shomali2022}
M. H. Fotovvat and Z. Shomali, A time-fractional dual-phase-lag framework to investigate transistors with TMTC channels (TiS$_3$, In$_4$Se$_3$) and size-dependent properties, Micro and nanostructures, 168, 207304, 2022.

\bibitem{Mazumder2001}
S. Mazumder and A. Majumdar, Monte carlo study of phonon transport in solid thin films including dispersion and polarization, J. Heat Mass Transf, 123, 749, 2001.

\bibitem{Mittal2011}
A. Mittal, Monte-Carlo Study of Phonon Heat Conduction in Silicon Thin Films, Diss. The Ohio State University, (2009).

\bibitem{Shomali2017}
Z. Shomali, B. Pedar, J. Ghazanfarian, and A. Abbassi, Monte-Carlo Parallel Simulation of Phonon Transport for 3D Nano-Devices, International Journal of Thermal Sciences, 114, 139, 2017.

\bibitem{2Shomali2017}
Z. Shomali, J. Ghazanfarian, and A. Abbassi, 3-D Atomistic Investigation of Silicon MOSFETs, ICHMT Digital Library Online, Begel House Inc., 2017.

	\bibitem{Wong2011}
	B.T. Wong, M. Francoeur, M.P. Meng, A Monte Carlo simulation for phonon transport within silicon structures at nanoscales with heat generation, Int. J. Heat Mass Transf. 54 (9) (2011) 1825.
	
	\bibitem{Wong2014}
	B.T. Wong, The impact of internal polarized monochromatic acoustic phonon emission on heat dissipation at nanoscale, Int. Commun. Heat Mass 53 (2014) 87.
	
	
		\end{thebibliography}
		\end{document}